\documentclass[pra,twocolumn,amsmath,amssymb,superscriptaddress]{revtex4-2}

\usepackage{epsfig,amsmath}
\usepackage{subfigure}
\usepackage{graphicx}
\usepackage{dcolumn}
\usepackage{stmaryrd}
\usepackage{mathrsfs}
\usepackage{pifont}
\usepackage{amsthm}
\usepackage{amssymb}
\usepackage{bm}
\usepackage{latexsym}
\usepackage[colorlinks=true,linkcolor=blue,citecolor=blue]{hyperref}
\usepackage{color}
\usepackage{epstopdf}

\begin{document}
\title{Kerr-magnon-assisted asymptotic stationary photon-phonon squeezing}
\author{Shi-fan Qi}
\email{Email address: qishifan@hebtu.edu.cn}
\affiliation{College of Physics and Hebei Key Laboratory of Photophysics Research and Application, Hebei Normal University, Shijiazhuang 050024, China}
\author{Jun Jing}
\email{Email address: jingjun@zju.edu.cn}
\affiliation{School of Physics, Zhejiang University, Hangzhou 310027, Zhejiang, China}

\begin{abstract}
Bosonic two-mode squeezed states are paradigmatic entangled states in continuous variable systems, which have broad applications in quantum information processing. In this work, we propose a photon-phonon squeezing protocol assisted by a Kerr magnon within a hybrid cavity magnomechanical system. We construct an effective Hamiltonian that accounts for photon-phonon squeezing through strong photon-magnon interaction and modulation over the driving on the photon mode. The effective Hamiltonian can be confirmed by the diagonalization of the system's Liouvillian superoperator. With the effective Hamiltonian and quantum Langevin equation, we provide a rigorous theoretical solution for the dynamical process of squeezing generation. Our finding indicates that the asymptotic stationary squeezing can be obtained by optimizing the squeezing quadrature operator, even when the covariance matrix of the system still varies with time. This squeezing level can exceed the maximum value under stable conditions. Moreover, our analysis reveals that a proper Kerr nonlinearity of the magnon can further promote the squeezing generation. Our work provides an extendable framework for generating squeezed states of two Gaussian modes with indirect coupling.
\end{abstract}

\maketitle
\section{Introduction}
Hybrid quantum systems consisting of collective spin excitations in ferromagnetic crystals have recently attracted intensive attention~\cite{cavitymagnonics,quantummagnon,bistabilitymagnon,yigcavity,gatemagnon,magnonqubit,qubitmagnon,kerrmagnoncav}. They offer promising avenues for advancements in quantum computing~\cite{quantumcomputing}, quantum communication~\cite{quantumcommunication}, and quantum sensing~\cite{quantumsense}. Similar to cavity-QED~\cite{circuit} and cavity optomechanics~\cite{optcavity}, cavity magnomechanics~\cite{magnoncavity,kerrcavitymagnon,BackactionMag,cavitymagnomechan} developed rapidly as an alternative candidate for quantum information processing in both theoretical and experimental aspects. In particular, a cavity magnomechanical system comprises a single-crystal yttrium iron garnet (YIG) sphere inside a microwave cavity. The magnon mode arises from excitations of the collective angular momentum within the magnetic-material sphere. It couples with the cavity photon through magnetic dipole interaction and the sphere deformation phonon mode via magnetostrictive force. Typical applications based on such hybrid system include quantum entanglement~\cite{mppentang,mppentangle} and steering~\cite{steermagnon,genuinesteer}, quantum squeezed states~\cite{squeezemag,squeeze,strongsqueeze}, and quantum memory~\cite{adiabaticpmp,intermagnon,acceleradia}.  

Two-mode squeezed states (TMSS), also named Einstein-Podolsky-Rosen (EPR) states, are crucial in quantum computation~\cite{quancomcon}, information~\cite{quaninfcon}, teleportation~\cite{quantelep}, and metrology~\cite{quantmetro}. Bosonic TMSS can be generated by mixing two single-mode squeezed states on a beam splitter~\cite{convariable} or via a nonlinear interaction~\cite{twocolor,twinlaser} such as spontaneous parametric down conversion~\cite{stablethreshold}. A nondegenerate optical parametric amplifier is often used to generate optical TMSS~\cite{quanphasenpo,realEPR,observasquee}. Cavity optomechanics~\cite{optcavity} provides an alternative model for creating optical~\cite{reservoiroptome,robustopto,doubleoptosys} or mechanical~\cite{twomechansquee,thermalsquee} TMSS. In addition, TMSS are well established in thermal gases~\cite{macroscopic}, Bose-Einstein condensates of ultracold atoms~\cite{homodyne,strongobserspin,probespin,spinBES}, spin ensembles in cavities~\cite{phasemeasure,manipulatinggrowth,powerlawspin,nongaussian}, antiferromagnet magnons~\cite{antiferromagnetic}, and superconducting circuits based on Josephson junctions~\cite{travelingwave,surfacephonon}.

In this work, we propose an approach to generate photon-phonon TMSS in the cavity magnomechanics~\cite{magnoncavity,kerrcavitymagnon,BackactionMag,cavitymagnomechan} on account of the fundamental interest in a level-resolved process. The squeezing generation is governed by an effective Hamiltonian that describes photon-phonon squeezing interaction, assisted by exploiting strong or even ultrastrong magnon-photon and magnon-phonon interactions. To confirm the validity of the effective Hamiltonian, we employ the diagonalization of the Liouvillian superoperator of the whole system. This approach can effectively address the squeezing Hamiltonian that does not conserve the whole excitations. It is thus distinct from the previous method~\cite{intermagnon,oneexcite} involving a standard numerical diagonalization of the system Hamiltonian. Additionally, our approach can be extended to arbitrary bosonic systems, such as cavity optomechanics~\cite{optcavity} and cavity optomagnomechanics~\cite{cavityoptomag}, to evaluate the two-mode squeezing induced by virtual processes.

Such a two-mode squeezing naturally leads to entanglement without reservoir engineering. Under the constraint of the stable condition, the squeezing level cannot go beyond $3$ ${\rm dB}$ below the vacuum limit~\cite{reservoiroptome}. However, by examining the system's dynamics within the open-quantum-system framework, we find that the stability of the Gaussian system, i.e., the covariance matrix (CM) becomes invariant when $t\to\infty$, is a sufficient but not necessary condition for the stationary generation of TMSS. We find that an asymptotic stationary TMSS can be obtained in unstable evolutions, displaying an enhanced squeezing level exceeding the steady limit. Environmental noises alter the optimized quadrature operator of two-mode squeezing while simultaneously stabilizing TMSS in an asymptotic way. The coupling between two modes can change the squeezing level, but it does not influence the squeezing stationarity, even if it is beyond the stability threshold~\cite{stablethreshold} of the CM. Through analysis of the system's dynamics under noise, we found that strong coupling essentially implies strong and stationary quantum entanglement.

The rest of this work is organized as follows. In Sec.~\ref{Secmodel}, we introduce a hybrid cavity magnomechanical system and provide an effective Hamiltonian for photon-phonon squeezing mediated by the magnon. In Sec.~\ref{Seceff}, we confirm the effective Hamiltonian by comparing the effective coupling strength and energy shift with numerical results obtained by diagonalization of the system's Liouvillian superoperator. Section~\ref{Secppsquee} phenomenologically analyzes the generation process of the photon-phonon TMSS by the quantum Langevin equation. We find that the asymptotic stationary two-mode squeezing can be obtained even in an unstable dynamic regime. Finally, we discuss the experimental feasibility and summarize the work in Sec.~\ref{Secexpercon}. 

\section{Model and the effective Hamiltonian}\label{Secmodel}
\begin{figure}[htbp]
\centering
\includegraphics[width=0.45\textwidth]{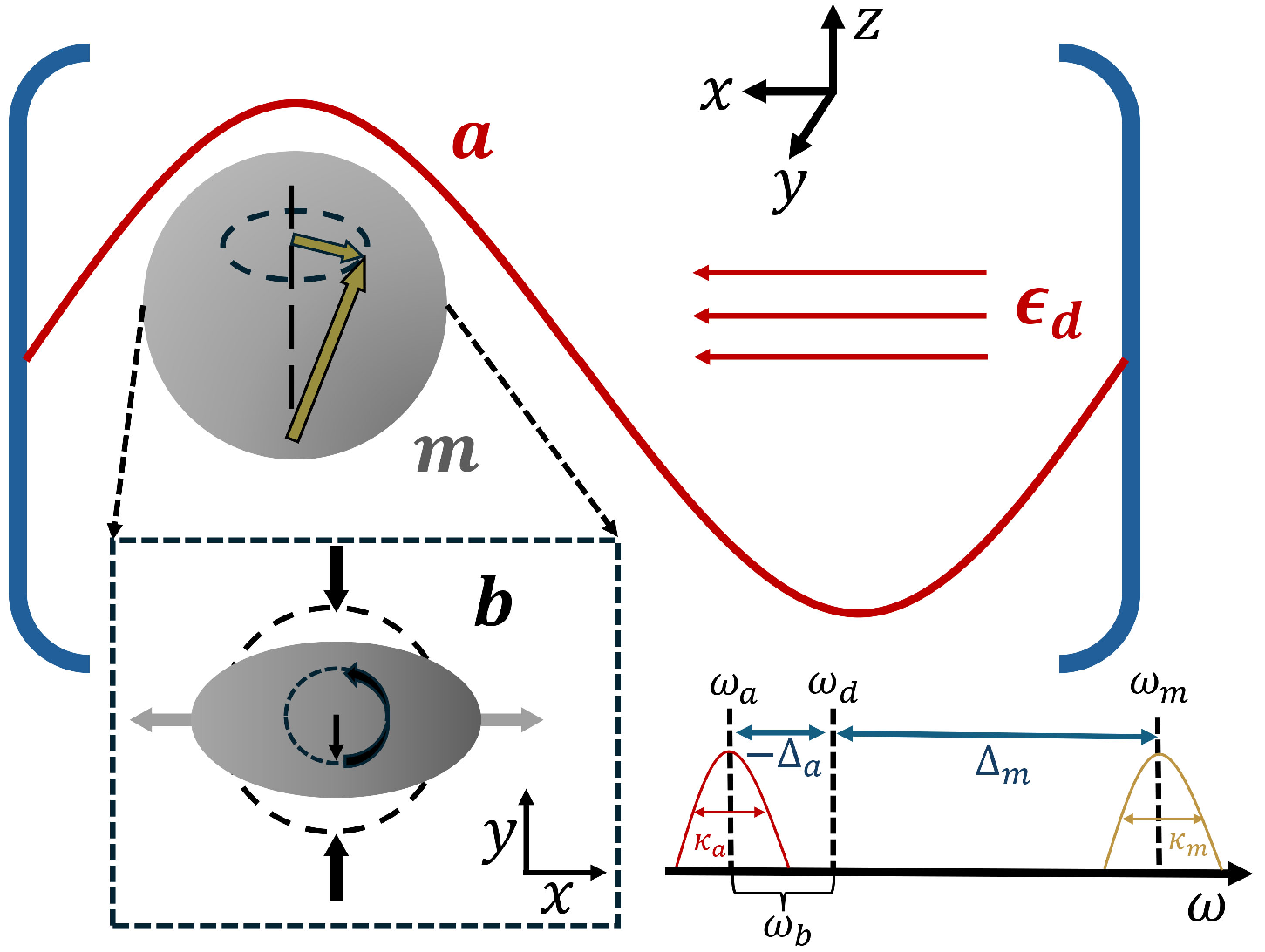}
\caption{Schematic diagram: a YIG sphere is placed inside a microwave cavity near the maximum magnetic field of the cavity mode, which establishes the magnon-photon coupling along the $z$ axis. The photon mode is driven by a microwave source along the $x$ axis (with a Rabi frequency $\epsilon_d$). The inset shows how the dynamic magnetization of a magnon (vertical black arrows) causes the deformation (compression along the $y$ direction) of the YIG sphere (and vice versa). Frequencies and linewidths of the system adopted to generate photon-phonon TMSS are shown in the bottom right corner.}\label{diagram}
\end{figure}

Consider a hybrid cavity magnomechanical system as shown in Fig.~\ref{diagram}, where a YIG sphere is inserted into a three-dimensional microwave cavity. The system is constituted by the microwave-mode photons, the magnons provided by the YIG sphere, and the vibrational modes (phonons) of the sphere. The magnons are coupled to photons via the Zeeman interaction and to phonons by the magnetostrictive interaction. The hybrid system has been experimentally realized in recent works~\cite{magnoncavity,kerrcavitymagnon,BackactionMag}. The full system Hamiltonian thus reads ($\hbar\equiv 1$)
\begin{equation}\label{Hamsysori}
\begin{aligned}
H_s&=\omega_aa^\dag a+\omega_mm^\dag m-K_mm^\dag mm^\dag m+\omega_bb^\dag b\\
&+g_{ma}(a^\dag m+am^\dag)+g_{mb}m^\dag m(b+b^\dag)\\
&+\epsilon_d(a^\dag e^{-i\omega_dt}+ae^{i\omega_dt}),
\end{aligned}
\end{equation}
where $a(a^\dag)$, $m(m^\dag)$, and $b(b^\dag)$ are the annihilation (creation) operators of the photon mode, the magnon of the ground Kittel mode~\cite{cavitymagnonics}, and the phonon mode with transition frequencies $\omega_a$, $\omega_m$, and $\omega_b$, respectively. The magnon-mode frequency is determined by $\omega_m=\gamma h$, where $\gamma$ is the gyromagnetic ratio and $h$ is the external bias magnetic field. Thus, it can be appropriately tuned by the external magnetic field. $K_m$ is the nonlinear coefficient for the Kerr effect due to the ensued magnetocrystalline anisotropy, which can be either positive or negative by adjusting the crystallographic axis of the YIG sphere along the bias magnetic field and is inversely proportional to the volume of the YIG sphere. The Kerr effect cannot be neglected when the magnon excitation number is sufficiently large~\cite{kerrcavitymagnon,kerrmagnonspin,kerrmagnoncav,bistabilitymagnon}. $g_{ma}$ is the photon-magnon coupling strength, entering into the strong-coupling regime. The single-magnon magnomechanical coupling strength $g_{mb}$ is typically small, considering the large frequency mismatch between the magnon and the phonon modes, yet it can be compensated by a strong drive. The last term in Eq.~\eqref{Hamsysori} describes the external driving of the photon mode, where $\epsilon_d$ is the Rabi frequency and $\omega_d$ is the driving frequency.

The magnon mode under strong driving is assumed to have a large expectation value $|\langle m\rangle|\gg1$, which allows us to linearize the system dynamics. Following the standard linearization approach~\cite{optcavity,mppentang}, the full system Hamiltonian turns out to be
\begin{equation}\label{Hamfull}
\begin{aligned}
H&=\Delta_aa^\dag a+\Delta'_mm^\dag m+\omega_bb^\dag b+g\cosh r(am^\dag+a^\dag m)\\
&+g\sinh r(ame^{-i\theta}+a^\dag m^\dag e^{i\theta})\\
&+Ge^r(me^{-i\frac{\theta}{2}}+m^\dag e^{i\frac{\theta}{2}})(b+b^\dag),
\end{aligned}
\end{equation}
where $\Delta'_m=\Delta_m/\cosh(2r)$ is the modified magnon detuning, $\Delta_m=\omega_m-\omega_d-2|K|$ and $K=K_m\langle m\rangle^2$ is the driving-enhanced Kerr parameter. $r$ is the squeezing parameter induced by the linearization of the Kerr effects and $\tanh(2r)=2|K|/\Delta_m$. $g=g_{ma}$ for simplicity, $G=g_{mb}|\langle m\rangle|$ is the driving-enhanced magnomechanical coupling strength. $\theta$ is a phase associated with $\langle m\rangle=|\langle m\rangle|e^{i\theta/2}$. The details can be found in Appendix~\ref{appalinearHam}.

At the large detuning regime, i.e., $g\cosh r, g\sinh r, Ge^r \ll |\Delta'_m-\omega_b|, |\Delta'_m-\Delta_a|$, and under the near-resonant condition $\Delta_a=-\omega_b+\delta$, we can extract an effective Hamiltonian describing the photon-phonon squeezing by perturbative theory~\cite{oneexcite,nonlinear}. The effective Hamiltonian is found to be
\begin{equation}\label{Heff}
H_{\rm eff}=g_{\rm eff}(e^{i\frac{\theta}{2}}a^\dag b^\dag+e^{-i\frac{\theta}{2}}ab),
\end{equation}
where the effective coupling strength is
\begin{equation}\label{geff}
g_{\rm eff}=gG\cosh(2r)\frac{\omega_b\cosh(2r)-\Delta_m e^{2r}}{\Delta^2_m-\omega^2_b\cosh^2(2r)},
\end{equation}
and the energy shift is
\begin{equation}\label{deltaeff}
\delta=\frac{2G^2\Delta_m e^{2r}\cosh(2r)+g^2(\Delta_m-\omega_b)\cosh^2(2r)}{\Delta^2_m-\omega^2_b\cosh^2(2r)}.
\end{equation}
The details can be found in Appendix~\ref{appaeffHam}. The phase $\theta$ specifies the squeezing quadrature operator but does not affect the squeezing level. Therefore, we set $\theta=\pi$ in the following for simplicity and with no loss of generality. The corresponding squeezing operators can be written as
\begin{equation}\label{XY}
X(t)=\frac{1}{\sqrt{2}}[X_a(t)+X_b(t)],
\end{equation}
where $X_o=(o+o^\dag)/\sqrt{2}, o=a,b$. When the initial state is a vacuum state, its variance turns into $\Delta X(t)=\langle X^2(t)\rangle-\langle X(t)\rangle^2=e^{2g_{\rm eff}t}/2$ under the time-evolution of Hamiltonian~\eqref{Heff}~\cite{stablethreshold,thermalsquee}. Obviously, the quadrature operator $X$ is squeezed, i.e., $\Delta X(t)<1/2$, when $g_{\rm eff}<0$. 

\section{The application range of the effective Hamiltonian}\label{Seceff}
\begin{figure}[htbp]
\centering
\includegraphics[width=0.48\textwidth]{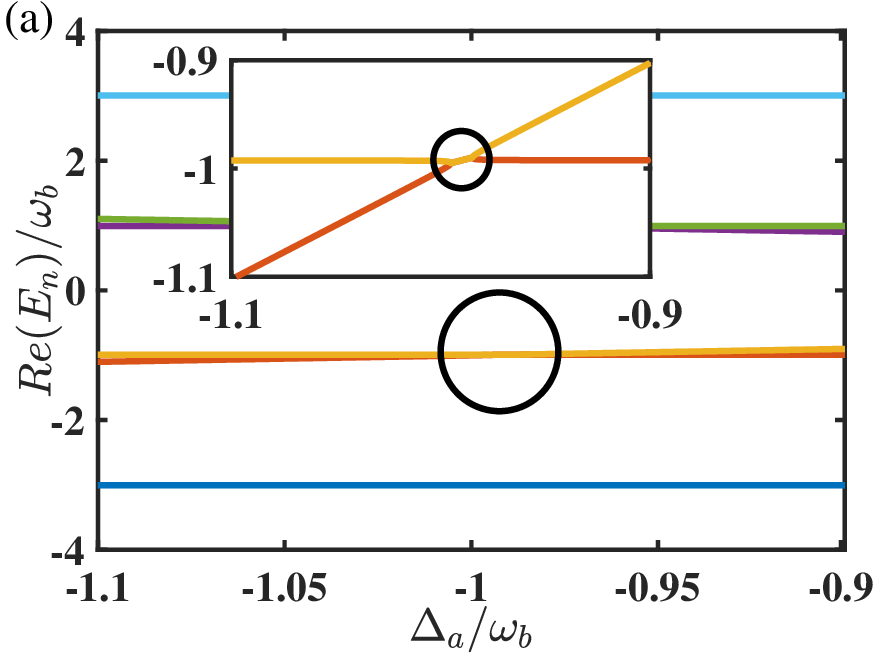}
\includegraphics[width=0.48\textwidth]{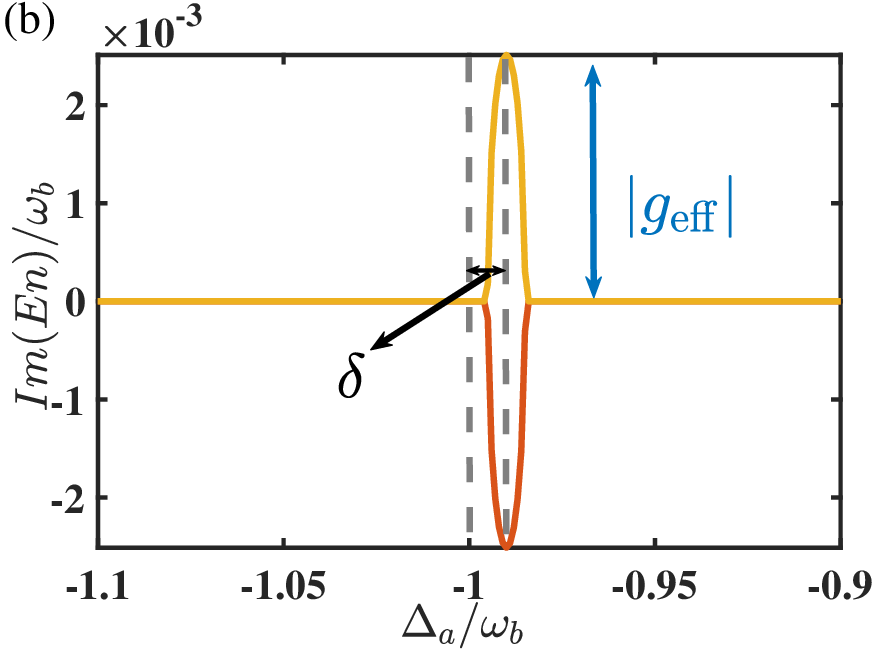}
\caption{(a) All six real parts of the normalized eigenvalues of the Liouvillian superoperator are depicted as a function of the detuning frequency $\Delta_a/\omega_b$. (b) Two relevant imaginary parts of the normalized eigenvalues are depicted as a function of the detuning frequency $\Delta_a/\omega_b$. The parameters used are $\Delta_m=3\omega_b$, $g=G=0.1\omega_b$, and $r=0$. }\label{eigenLiouvillian}
\end{figure}

\begin{figure}[htbp]
\centering
\includegraphics[width=0.235\textwidth]{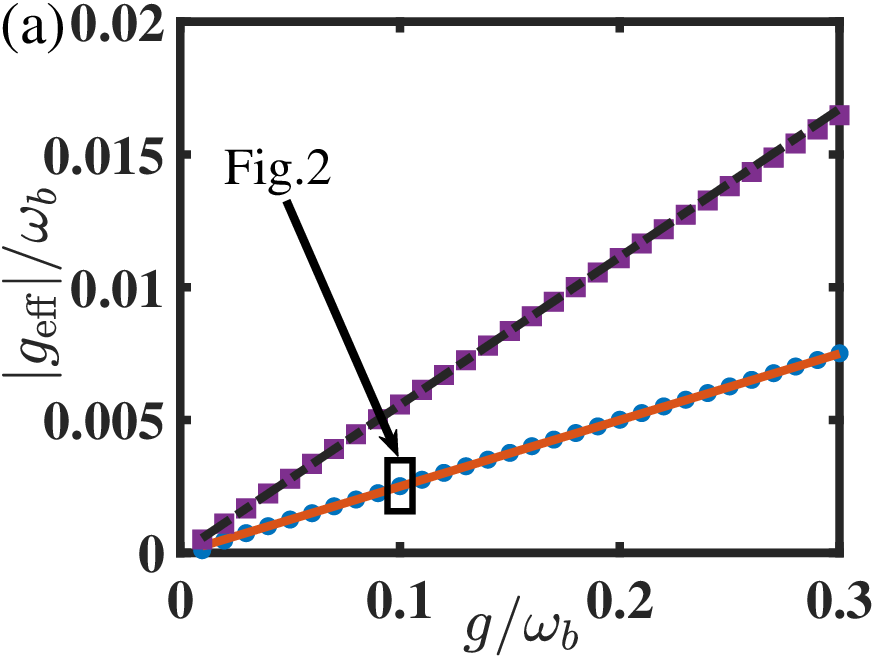}
\includegraphics[width=0.235\textwidth]{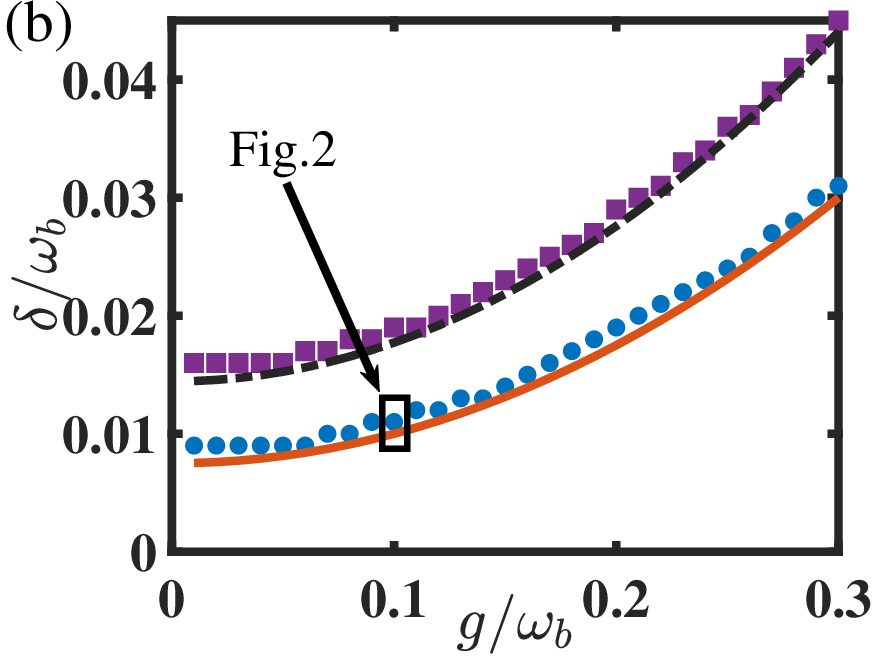}
\includegraphics[width=0.235\textwidth]{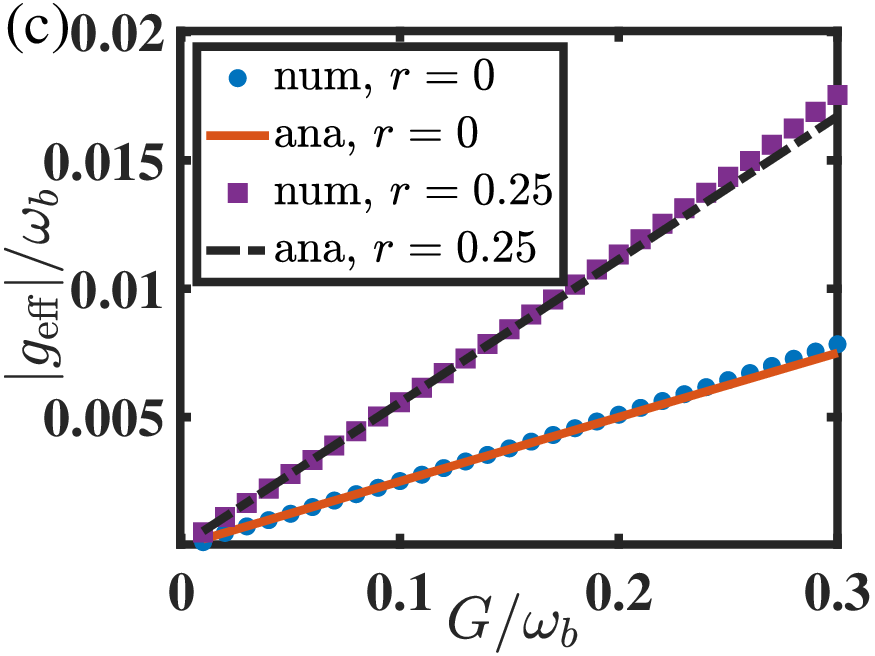}
\includegraphics[width=0.235\textwidth]{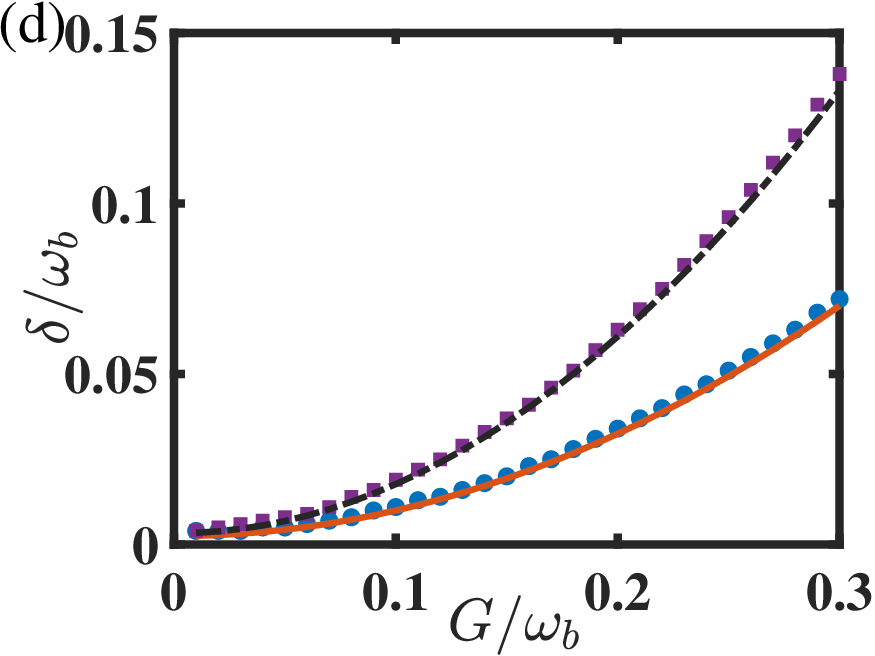}
\includegraphics[width=0.235\textwidth]{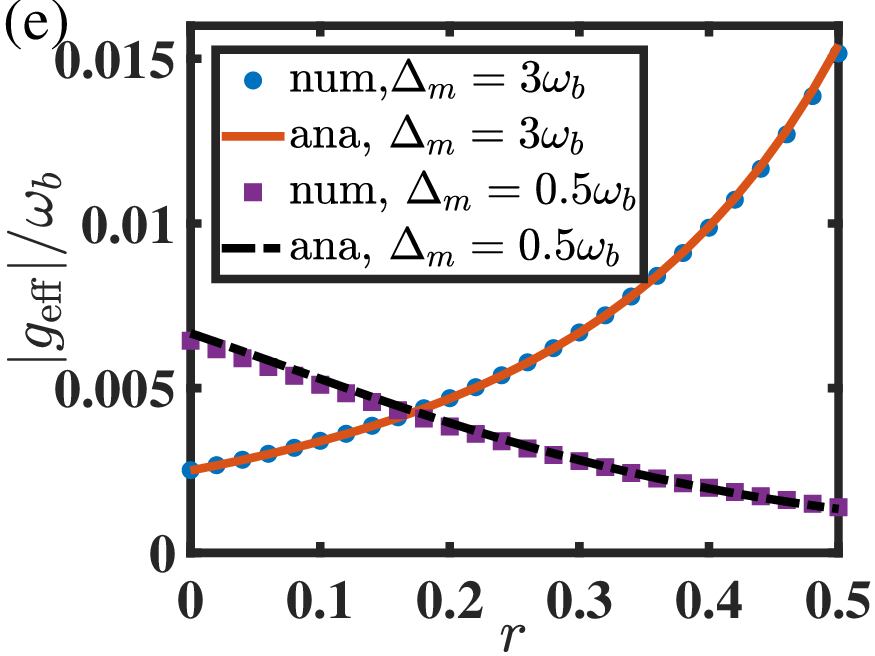}
\includegraphics[width=0.235\textwidth]{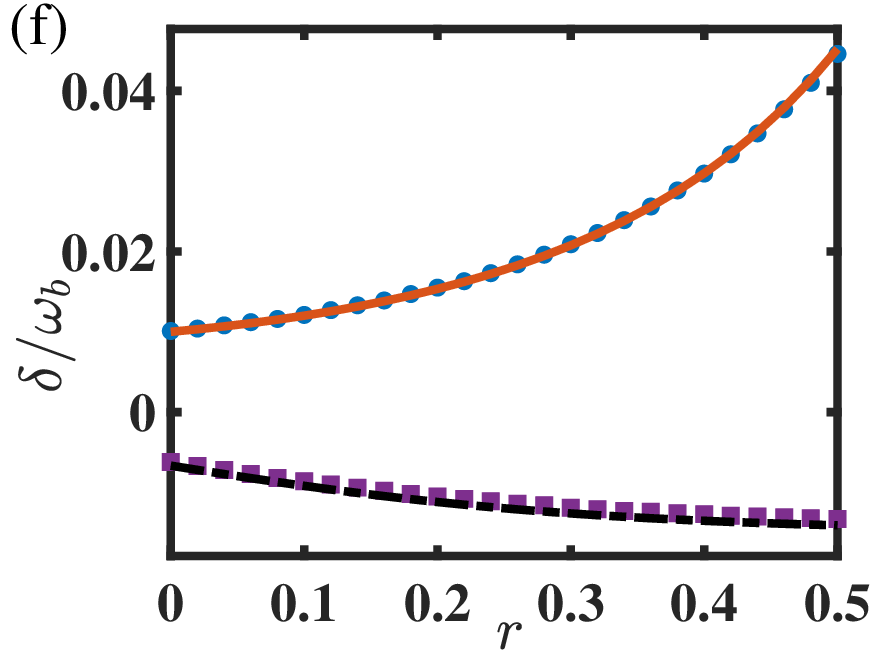}
\caption{[(a), (c), (e)] Comparison between the numerically calculated normalized effective coupling strength $|g_{\rm eff}|/\omega_b$ (points) and the corresponding analytical results (lines) in Eq.~\eqref{geff} as a function of $g/\omega_b$, $G/\omega_b$, and squeezing parameter $r$, respectively. [(b), (d), (f)] Comparison between the numerically calculated normalized energy shift $\delta/\omega_b$ (points) and the corresponding analytical results (lines) in Eq.~\eqref{deltaeff} as a function of $g/\omega_b$, $G/\omega_b$, and squeezing parameter $r$, respectively. Here, $G=0.1\omega_b,\Delta_m=3\omega_b$ for panels (a) and (b), $g=0.1\omega_b,\Delta_m=3\omega_b$ for panels (c) and (d), and $g=G=0.1\omega_b$ for panels (e) and (f). The legends for panels (a)-(d) are combined in panel (c), while the legends for panels (e) and (f) are combined in panel (e).}\label{effgdelta}
\end{figure}

In this section, we check the applicability range of the effective Hamiltonian in Eq.~\eqref{Heff} in terms of the coupling strengths and squeezing parameters. In previous works~\cite{oneexcite,intermagnon}, diagonalizing the full system Hamiltonian in a truncated finite-dimensional Hilbert space is employed to confirm the effective Hamiltonian constructed by virtual transitions. It is essential to observe a desired avoided level crossing between two eigenstates in the eigenenergies diagram, and the energy splitting at the avoided level crossing point precisely equals twice the effective coupling strength. However, the two-mode squeezing effective Hamiltonian~\eqref{Heff} is not conserved in the excitation number, as shown by the non-commutativity $[H_{\rm eff},\hat{N}]\neq0$ with $\hat{N}=a^\dag a+b^\dag b$ the excitation-number operator. Thus, the effective Hamiltonian~\eqref{Heff} cannot be rigorously diagonalized within an appropriate truncated Hilbert space. Additionally, an explicit avoided-level crossing between two eigenstates does not appear for the two-mode squeezing Hamiltonian. Consequently, we propose a distinct method to validate the two-mode squeezing Hamiltonian in Eq.~\eqref{Heff}.

Under the evolution of the full system Hamiltonian in Eq.~\eqref{Hamfull}, the time-evolved quadrature operators in the Heisenberg picture can be written as 
\begin{equation}\label{utHeisenberg}
\dot{u}(t)=i[H,u(t)]=i\mathcal{L}u(t),
\end{equation}
where $u(t)=[X_a(t), Y_a(t), X_b(t), Y_b(t), X_m(t), Y_m(t)]^T$ is the vector of quadrature operators, and $X_o=(o+o^\dag)/\sqrt{2}, Y_o=(o-o^\dag)/i\sqrt{2}, o=a,b,m$. $\mathcal{L}$ represents the Liouvillian superoperator, 
\begin{equation}\label{Liouvilliansuper}
\begin{aligned}
\mathcal{L}=-i\begin{bmatrix}
0 &\Delta_a& 0 & 0&0&g_+\\
-\Delta_a&0&0&0&g_-&0\\
0&0&0&\omega_b&0&0\\
0&0&-\omega_b&0&0&-G_r\\
0&g_+&G_r&0&0&\Delta'_m\\
g_-&0&0&0&-\Delta'_m&0 
\end{bmatrix},
\end{aligned}
\end{equation}
where $g_{\pm}=g\sinh r\pm g\cosh r$ and $G_r=2Ge^{r}$. The Heisenberg equation in Eq.~\eqref{utHeisenberg} can be regarded as a discrete Schr\"{o}dinger equation, where $u(t)$ is conceptualized as an effective operator wave function~\cite{delocalization}. The superoperator $\mathcal{L}$ then can be analogously regarded as the full system Hamiltonian, and its diagonalization values are the system's eigenvalues.

We now analyze the distinct phenomenon observed in the energy diagram of the Liouvillian superoperator at two-mode squeezing. Rotating the effective Hamiltonian~\eqref{Heff} with $\theta=\pi$ into the lab frame, it becomes 
\begin{equation}
H_{\rm ab}=\Delta_a a^\dag a+\omega_bb^\dag b+g_{\rm eff}(ia^\dag b^\dag-iab).
\end{equation}
The corresponding Heisenberg equation is 
\begin{equation}\label{lioueff}
\dot{u}^{\rm eff} (t)=i[H_{\rm ab},u^{\rm eff}(t)]=i\mathcal{L}_{\rm ab}u^{\rm eff}(t),
\end{equation}
where $u^{\rm eff}(t)=[X_a(t), Y_a(t), X_b(t), Y_b(t)]^T$. Four eigenvalues of the Liouvbilian superoperator $\mathcal{L}_{\rm ab}$ can be derived as 
\begin{equation}
\begin{aligned}
E_{\pm}&=\frac{\omega_b-\Delta_a\pm\sqrt{(\omega_b+\Delta_a)^2-4g_{\rm eff}^2}}{2},\\
E'_{\pm}&=-E_{\mp}.
\end{aligned}
\end{equation} 
The real parts of the eigenvalues $E_{\pm}$ ($E'_{\pm}$) converge, while the imaginary parts split as the detuning $\Delta_a$ gradually approaches $-\omega_b$. Until $\Delta_a=-\omega_b$, the real parts of  $E_{\pm}$ ($E'_{\pm}$) become identical, while the imaginary parts reach their extreme values of $\pm g_{\rm eff}$. Then, in the energy-level diagram of the whole superoperator $\mathcal{L}$ as a function of $\Delta_a$, one can demonstrate the two-mode squeezing interaction through the level attractions of the real parts and the maximal splittings of the imaginary parts. 

We plot the energy levels (all six real and two relevant imaginary parts) in Figs.~\ref{eigenLiouvillian}(a) and~\ref{eigenLiouvillian}(b), where the eigenvalues $\{E_n\}$ are obtained by the standard numerical diagonalization on the whole superoperator $\mathcal{L}$ in Eq.~\eqref{Liouvilliansuper}. Figure~\ref{eigenLiouvillian}(a) shows the real parts of all six eigenvalues. The light-blue and dark-blue lines describe the energies of the magnon and are not relevant to photon-phonon squeezing. For the other four eigenvalues, two level attractions appear simultaneously as the detuning frequency of photon $\Delta_a$ approaches (but does not exactly equal) the opposite frequency of phonon $-\omega_b$. This level attraction is highlighted by a dark circle, and the inset further emphasizes it. The imaginary parts of the two relevant eigenvalues (red and orange lines) are illustrated in Fig.~\ref{eigenLiouvillian}(b). As the real parts of the two eigenvalues gradually converge, their imaginary parts progressively increase, reaching a maximum absolute value $|g_{\rm eff}|$ at $\Delta_a=-\omega_b+\delta$. The shift $\delta$ is induced by the mutual interaction between the photon (phonon) and the magnon. 

The maximal splitting $|g_{\rm eff}|$ of the imaginary parts of the two eigenvalues [see Fig.~\ref{eigenLiouvillian}(b)] is presented in Figs.~\ref{effgdelta}(a) and~\ref{effgdelta}(c) as a function of the original coupling strengths $g$ and $G$ in Eq.~\eqref{Hamfull}, respectively. The analytical result in Eq.~\eqref{geff} is compared to the numerical simulation over the superoperator $\mathcal{L}$ in Eq.~\eqref{Liouvilliansuper}. Blue dots and purple squares represent the numerical results at squeezing parameters $r=0$ and $r=0.25$, respectively. The red solid and black dashed lines are the analytical results at $r=0$ and $r=0.25$, respectively. In Fig.~\ref{effgdelta}(a), the analytical $g_{\rm eff}$ do match well with their numerical results for the coupling strength $g\le0.3\omega_b$, regardless of whether $r$ is $0$ or $0.25$. The valid range has entered the ultrastrong coupling regime, $g/\omega_b\ge0.1$~\cite{Ultrastrong}. The value distinguished by the black box corresponds to Fig.~\ref{eigenLiouvillian}(b). In Fig.~\ref{effgdelta}(c), $g_{\rm eff}$ is valid at the range of $G/\omega_b\le 0.3$ when $r=0$. As $r$ increases to $0.25$, the valid range reduces to $G/\omega_b\le 0.24$. Similarly, the energy shift $\delta$ in Eq.~\eqref{deltaeff} can also be justified by Figs.~\ref{effgdelta}(b) and~\ref{effgdelta}(d). In Fig.~\ref{effgdelta}(b), the analytical $\delta$ shows a slight deviation from the numerical results, but this deviation gradually decreases as $g$ and $r$ increase. In Fig.~\ref{effgdelta}(d), it is found that the energy shift $\delta$ is valid at $G/\omega_b\le 0.3$, whatever the parameter $r$ is $0$ or $0.25$. 

Similar results about $g_{\rm eff}$ and $\delta$ are plotted as a function of the squeezing parameter $r$ in Figs.~\ref{effgdelta}(e) and (f), respectively. It is evident that both the analytical results for $g_{\rm eff}$ and $\delta$ match well with the numerical results at $\Delta_m=3\omega_b$. The effective coupling $g_{\rm eff}$ is significantly amplified as $r$ increases, which can improve the photon-phonon squeezing level discussed later. However, at $\Delta_m=0.5\omega_b$, there exists a slight distinction between the numerical and analytical results for both $g_{\rm eff}$ or $\delta$. This distinction is mainly attributed to the constraint of the parameter setting, which yields that the detuning between magnon and phonon, $|\Delta_m/\cosh(2r)-\omega_b|$, is not sufficiently larger than the coupling strengths. Furthermore, $g_{\rm eff}$ decreases as $r$ increases, leading to a low two-mode squeezing level.

\section{Photon-phonon squeezing}\label{Secppsquee}
\begin{figure}[htbp]
\centering
\includegraphics[width=0.48\textwidth]{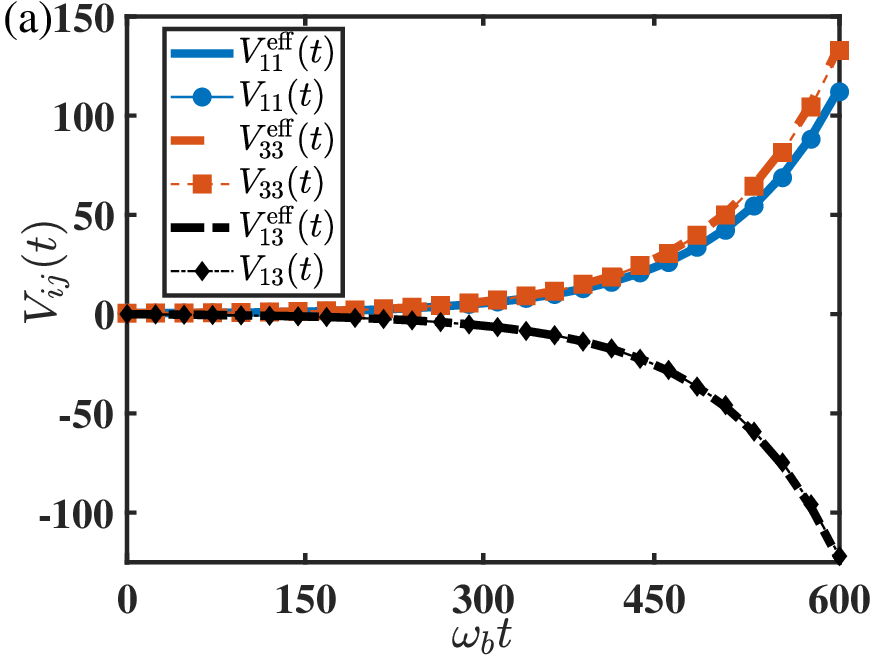}
\includegraphics[width=0.48\textwidth]{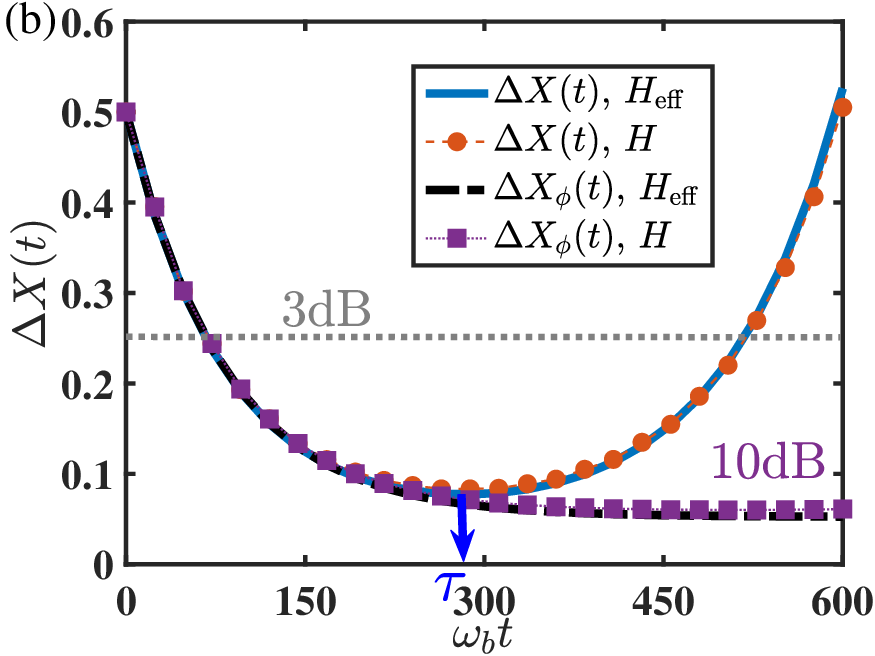}
\caption{(a) Dynamics of the CM elements using the effective Hamiltonian~\eqref{Heff} or the full system Hamiltonian~\eqref{Hamfull}. (b) Dynamics of the $\Delta X(t)$ and $\Delta X_\phi(t)$ with the effective Hamiltonian~\eqref{Heff} or the full system Hamiltonian~\eqref{Hamfull}. The parameters are set as $\Delta_m=3\omega_b$, $g=G=0.1\omega_b$, $r=0.25$, $\kappa_b=10^{-5}\omega_b,\kappa_a=100\kappa_b,\kappa_m=10\kappa_a$, and the thermal numbers $N_a=N_m=0, N_b=10$.}\label{pptotsqueeze}
\end{figure}

Using the effective Hamiltonian in Eq.~\eqref{Heff}, one can generate naturally and directly the photon-phonon TMSS. In this section, we take the open-quantum-system framework to discuss the dynamics of TMSS generation and explain why the asymptotic two-mode squeezing can be obtained even in an unstable dynamical evolution. Under the standard assumptions, i.e., Markovian approximation and structure-free environment at zero temperature, the dynamics of the quantum system are governed by the quantum Langevin equation (QLE), written in a matrix form
\begin{equation}\label{ulangevineff}
\dot{u}^{\rm eff}(t)=A_{\rm eff}u^{\rm eff}(t)+n^{\rm eff}(t),
\end{equation}
where $u^{\rm eff}(t)$ is the same as Eq.~\eqref{lioueff}. The transition matrix $A_{\rm eff}(t)$ is
\begin{equation}\label{Ateffmatrix}
\begin{aligned}
A_{\rm eff}=\begin{bmatrix}
-\kappa_a&0&g_{\rm eff}&0\\
0&-\kappa_a&0&-g_{\rm eff}\\
g_{\rm eff}&0&-\kappa_b&0\\
0&-g_{\rm eff}&0&-\kappa_b
\end{bmatrix}
\end{aligned}
\end{equation}
where $\kappa_a$ and $\kappa_b$ are the decay rates of the modes $a$ and $b$, respectively. $n^{\rm eff}(t)=[\sqrt{2\kappa_a}X^{in}_a(t), \sqrt{2\kappa_a}Y^{in}_a(t), \sqrt{2\kappa_b}X^{in}_b(t), \sqrt{2\kappa_b}Y^{in}_b(t) ]^T$ is the vector of Gaussian noise operators, and $X^{in}_o=(o_{in}+o^\dag_{in})/\sqrt{2}, Y^{in}_o=(o_{in}-o^\dag_{in})/i\sqrt{2}, o=a,b$. $a_{in}$ and $b_{in}$ are characterized by their covariance functions, $\langle o_{in}(t)o^\dag_{in}(t')\rangle=[N_o+1]\delta(t-t')$ and $\langle o^\dag_{in}(t)o_{in}(t')\rangle=N_o\delta(t-t')$, where $N_o$ is the mean population of mode $o$ at the thermal equilibrium state.

The input zero-mean quantum Gaussian noises yield the quantum state as a zero-mean Gaussian state, which can be completely characterized by a $4\times 4$ covariance matrix (CM) $V^{\rm eff}(t)$. By virtue of the QLE in Eq~\eqref{ulangevineff}, the dynamics of the CM $V^{\rm eff}(t)$ satisfies
\begin{equation}\label{Vijteff}
\dot{V}^{\rm eff}(t)=A_{\rm eff}V^{\rm eff}(t)+V^{\rm eff}(t)A_{\rm eff}^T+D^{\rm eff}.
\end{equation}
The elements of $V^{\rm eff}(t)$ are defined as 
\begin{equation}\label{Vijdefi}
V^{\rm eff}_{ij}(t)=\frac{\langle u^{\rm eff}_i(t)u^{\rm eff}_j(t)+u^{\rm eff}_j(t)u^{\rm eff}_i(t)\rangle}{2},\\
\end{equation}
where $u^{\rm eff}_i(t)$ is the $i$-term of $u^{\rm eff}(t)$ and $i=1,2,3,4$. $D^{\rm eff}=Diag[\kappa_a(2N_a+1),\kappa_a(2N_a+1),\kappa_b(2N_b+1),\kappa_b(2N_b+1)]$ is the diffusion matrix, which is defined through $D^{\rm eff}_{ij}(t)=\langle n^{\rm eff}_i(t)n^{\rm eff}_j(t)+n^{\rm eff}_j(t)n^{\rm eff}_i(t)\rangle/2$. In the stable condition, the CM is invariant under time evolution, i.e, $\dot{V}^{\rm eff}=0$ in Eq.~\eqref{Vijteff}, which requires $g^2_{\rm eff}<\kappa_a\kappa_b$.

Assume the photon and phonon are both in vacuum states initially, which can be realized by precooling them to their respective ground states~\cite{precooling}. Then the initial CM can be written as $V^{\rm eff}(0)=I_4/2$, $I_4$ is an identity matrix with four dimensions. Under this initial condition, the non-zero matrix elements in $V^{\rm eff}(t)$ can be solved as
\begin{equation}\label{V13t}
\begin{aligned}
V^{\rm eff}_{11}(t)&=C_+(1-\sin\varphi)e^{(\Omega-\kappa_a-\kappa_b)t}-C_0\cos\varphi e^{-(\kappa_a+\kappa_b)t}\\
&+C_-(1+\sin\varphi)e^{-(\Omega+\kappa_a+\kappa_b)t}+c_a;\\
V^{\rm eff}_{33}(t)&=C_+(1+\sin\varphi)e^{(\Omega-\kappa_a-\kappa_b)t}+C_0\cos\varphi e^{-(\kappa_a+\kappa_b)t}\\
&+C_-(1-\sin\varphi)e^{-(\Omega+\kappa_a+\kappa_b)t}+c_b;\\
V^{\rm eff}_{13}(t)&=C_+\cos\varphi e^{(\Omega-\kappa_a-\kappa_b)t}-C_0\sin\varphi e^{-(\kappa_a+\kappa_b)t}\\
&-C_-\cos\varphi e^{-(\Omega+\kappa_a+\kappa_b)t}+c,
\end{aligned}
\end{equation}
and $V^{\rm eff}_{22}(t)=V^{\rm eff}_{11}(t), V^{\rm eff}_{44}(t)=V^{\rm eff}_{33}(t), V^{\rm eff}_{24}(t)=-V^{\rm eff}_{13}(t)$. The parameters are defined as
\begin{equation}\label{constants}
\begin{aligned}
\Omega&=\sqrt{4g^2_{\rm eff}+(\kappa_a-\kappa_b)^2}, \quad \tan\varphi=\frac{\kappa_a-\kappa_b}{2g_{\rm eff}},\\
C_{\pm}&=\pm \frac{\kappa_+\mp\sin\varphi\kappa_-}{4[\Omega\mp(\kappa_a+\kappa_b)]}+\frac{1}{4},\quad C_0=\frac{\cos\varphi\kappa_-}{2(\kappa_a+\kappa_b)},\\
\kappa_\pm&=\kappa_a(2N_a+1)\pm\kappa_b(2N_b+1).
\end{aligned}
\end{equation}
And
\begin{equation}\label{steadyV}
\begin{aligned}
c_o&=N_o+\frac{1}{2}+\frac{g_{\rm eff}}{\kappa_o}c,\quad o=a,b\\
c&=\frac{g_{\rm eff}\kappa_a\kappa_b(N_a+N_b+1)}{(\kappa_a\kappa_b-g^2_{\rm eff})(\kappa_a+\kappa_b)},
\end{aligned}
\end{equation}
which are also the solutions of the matrix elements $V^{\rm eff}_{11}$, $V^{\rm eff}_{33}$, and $V^{\rm eff}_{13}$ in the stable regime, respectively, obtained by setting $\dot{V}^{\rm eff}=0$ in Eq.~\eqref{Vijteff}. These stable CM elements are the asymptotic values as $t\to \infty$.

With the CM definition in Eq.~\eqref{Vijdefi} and its solution in Eq.~\eqref{V13t}, the variance of the quadrature operator $X$ in Eq.~\eqref{XY} can be derived as
\begin{equation}\label{XYVeff}
\begin{aligned}
\Delta X(t)&=\frac{1}{2}[V^{\rm eff}_{11}(t)+V^{\rm eff}_{33}(t)+2V^{\rm eff}_{13}(t)]\\
&=(1+\cos\varphi)C_+e^{(\Omega-\kappa_a-\kappa_b)t}-\sin\varphi C_0e^{-(\kappa_a+\kappa_b)t}\\
&+(1-\cos\varphi)C_-e^{-(\Omega+\kappa_a+\kappa_b)t}+C,
\end{aligned}
\end{equation}
where 
\begin{equation}\label{constC}
\begin{aligned}
C&=\frac{1}{2}(N_a+N_b+1)\frac{\kappa_a\kappa_b(2g_{\rm eff}+\kappa_a+\kappa_b)}{(\kappa_a\kappa_b-g^2_{\rm eff})(\kappa_a+\kappa_b)}.
\end{aligned}
\end{equation}

In the stable regime, $g^2_{\rm eff}<\kappa_a\kappa_b$, one can demonstrate that the exponent factor $\Omega-\kappa_a-\kappa_b$ in Eq.~\eqref{XYVeff} is negative through the definition of $\Omega$ in Eq.~\eqref{constants}. That leads to $\Delta X(\infty)=C$, consistent with the result obtained by $\dot{V}^{\rm eff}(t)=0$. When the photon decay rate is larger than the phonon decay rate, i.e., $\kappa_a>\kappa_b$, the minimal value of $C$ can be obtained as
\begin{equation}\label{Cmin}
C_{\rm min}=\frac{1}{2}(N_a+N_b+1)\frac{\kappa_a}{\kappa_a+\kappa_b}
\end{equation}
at $g_{\rm eff}=-\kappa_b$. Even at the zero temperature, $N_a=N_b=0$, the minimum $C_{\rm min}>0.25$. The value $0.25$ corresponds to the upper bound of the squeezing level, $S=3$dB, under the stable condition. The squeezing level $S$ in the decibel unit is defined by $S=-10\log_{10}(\Delta X/\Delta X_{zp})$~\cite{unresolved}, where $\Delta X_{zp}=0.5$ is the standard fluctuation in the zero-point level. The decay rates satisfy $\kappa_a\gg\kappa_b$ in the recent cavity magnomechanical system~\cite{magnoncavity}, resulting in $C_{\rm min}\approx 0.5$ even when $N_a=N_b=0$. It implies that $X$ cannot be squeezed under stable conditions in this specific experimental platform.

In the unstable regime, $g^2_{\rm eff}>\kappa_a\kappa_b$, all the CM elements $V^{\rm eff}_{11}$, $V^{\rm eff}_{33}$, and $V^{\rm eff}_{13}$ in Eq.~\eqref{V13t} exhibit exponential divergence due to the exponential factor $\Omega-\kappa_a-\kappa_b>0$. These are clearly illustrated by their respective numerical results, shown by a blue-solid line, a red-dashed line, and a black dash-dotted line in Fig.~\ref{pptotsqueeze}(a). The corresponding variance $\Delta X(t)$ in Eq.~\eqref{XYVeff} is depicted by a blue solid line in Fig.~\ref{pptotsqueeze}(b). One can observe that it initially decreases until it reaches its minimum value $\Delta X(\tau)$, where the moment $\tau$ can be analytically determined by setting the derivation $ \dot{\Delta X}(\tau)=0$. After reaching its minimum, $\Delta X(t)$ increases exponentially, and $\Delta X(\infty)\to+\infty$. 

However, both the CM and the variance $\Delta X(t)$ instabilities do not imply nonstationary TMSS. To find a stationary TMSS with a higher squeezing level, we define a general two-mode squeezing operator $X_\phi=\cos\phi X_a+\sin\phi X_b$, where $\phi$ is an angle to optimize. With the CM elements~\eqref{V13t}, its variance $\Delta X_\phi=\langle X_\phi^2\rangle-\langle X_\phi\rangle^2$ can be described as
\begin{equation}\label{Deltaxeff}
\begin{aligned}
\Delta X_\phi(t)&=\cos^2\phi V^{\rm eff}_{11}(t)+\sin^2\phi V^{\rm eff}_{33}(t)+\sin(2\phi)V^{\rm eff}_{13}(t)\\
&=C_+(1-\sin\tilde{\varphi})e^{(\Omega-\kappa_a-\kappa_b)t}-C_0\cos\tilde{\varphi}e^{-(\kappa_a+\kappa_b)t}\\
&+C_-(1+\sin\tilde{\varphi})e^{-(\Omega+\kappa_a+\kappa_b)t}+C_\phi,\\
\end{aligned}
\end{equation}
where $\tilde{\varphi}=\varphi-2\phi$ and $C_\phi=\cos^2\phi c_a+\sin^2\phi c_b+\sin(2\phi) c$, $c_a, c_b, c$ are constants in Eq.~\eqref{steadyV}.

From Eq.~\eqref{Deltaxeff}, one can find that the exponential divergence term of $\Delta X_\phi(t)$ can be canceled at an optimized angle $\tilde{\varphi}=\pi/2$, i.e., the angle $\phi$ satisfies
\begin{equation}\label{phi}
\tan(2\phi)=-\cot(\varphi)=\frac{2g_{\rm eff}}{\kappa_b-\kappa_a}.
\end{equation}
Specifically, $\Delta X_\phi(t)=\Delta X(t)$ at $\kappa_a=\kappa_b$. Under this optimized angle, $\Delta X_\phi(t)$ turns into
\begin{equation}\label{Deltaxphit}
\Delta X_\phi(t)=\frac{1}{2}+2C_-e^{-(\Omega+\kappa_a+\kappa_b)t}-2C_-.
\end{equation}

It is evident that $\Delta X_\phi(0)=0.5$, corresponding to the standard fluctuation in the zero-point level~\cite{unresolved}. The condition $\Delta X_\phi(t)<\Delta X_\phi(0)$ signifies the occurrence of two-mode squeezing during the evolution~\cite{unresolved}. The value of $\Delta X_\phi(t)$ quantifies the level of the squeezing, a smaller $\Delta X_{\phi}(t)$ yielding a stronger squeezing. Equation~\eqref{Deltaxphit} also shows an asymptotic stationary squeezing over a long evolution, i.e., 
\begin{equation}\label{Deltaxphiinfty}
\Delta X_\phi(\infty)=\frac{\Omega\kappa_++(\kappa_a-\kappa_b)\kappa_-}{2\Omega(\Omega+\kappa_a+\kappa_b)}.
\end{equation}
Given the definition of $\Omega$ in Eq.~\eqref{constants}, it follows that $\Delta X_\phi(\infty)$ decreases as well as the two-mode squeezing enhances as $g_{\rm eff}$ increases.

It is a natural result of the system's evolution under the two-mode squeezing Hamiltonian~\eqref{Heff} in the open quantum system framework. The two-mode squeezing interaction generates the photon-phonon squeezing, gradually increasing the squeezing level over time. In contrast, Markovian noises reduce the squeezing as time progresses. Both factors constitute a competitive mechanism in the open-quantum-system framework, leading to an asymptotic stationary photon-phonon squeezing with an optimized squeezing operator.

In Fig.~\ref{pptotsqueeze} (b), we plot $\Delta X_\phi(t)$ using the effective Hamiltonian by blue dash-dotted line. After a long time evolution $\omega_b t\ge 450\approx 1.5\tau$, it tends to stabilize a certain value of $0.05$, and the corresponding squeezing level $S$ is about $10$dB below vacuum fluctuation, which is larger than the upper bound $3$dB in the stable dynamic condition.

\begin{figure}[htbp]
\centering
\includegraphics[width=0.235\textwidth]{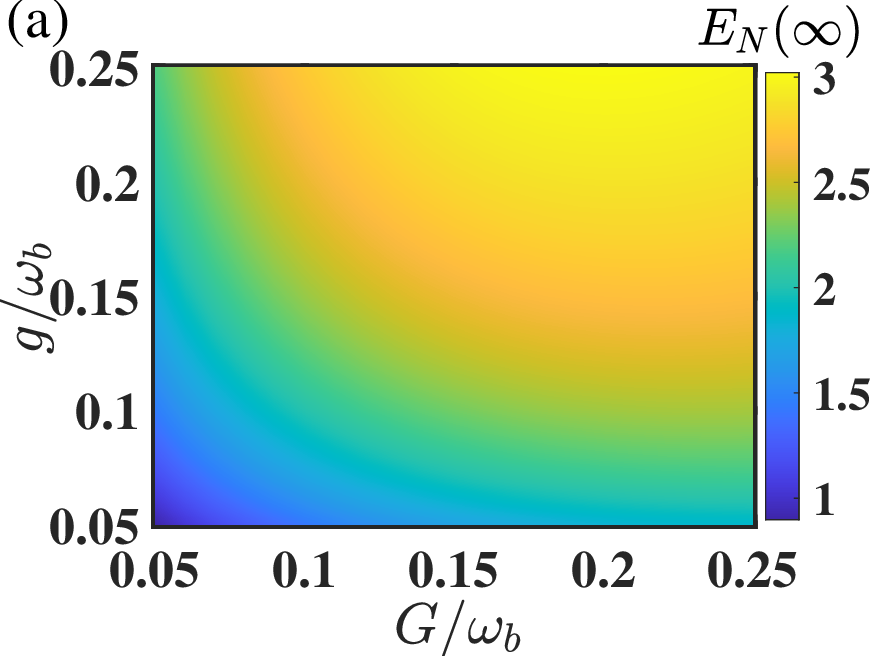}
\includegraphics[width=0.235\textwidth]{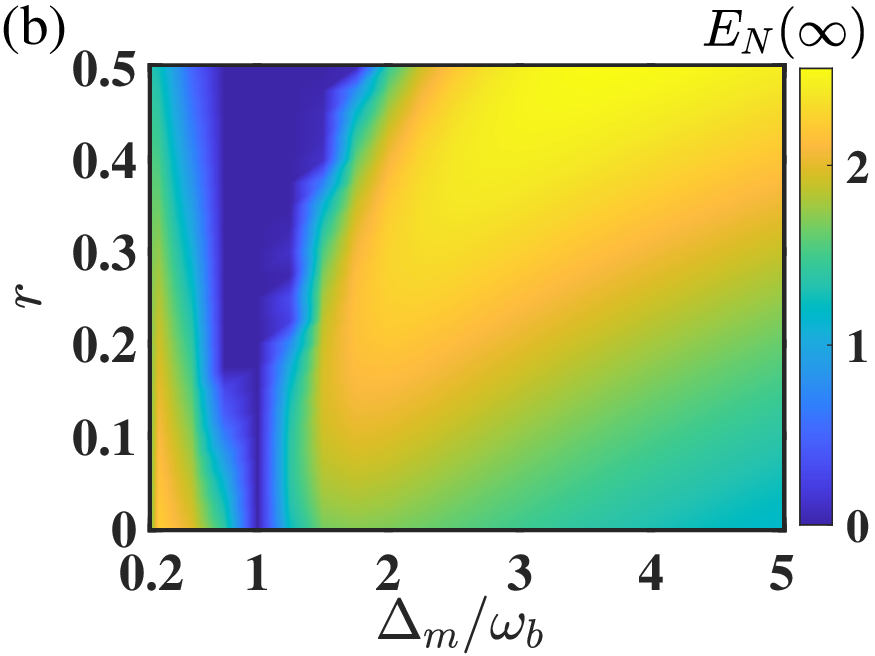}
\caption{(a) LN $E_N(\infty)$ in the coupling strengths $g$ and $G$ parameter space. Here, $\Delta_m=3\omega_b$ and $r=0.25$. (b) $E_N(\infty)$ in the parameter space spanned by magnon detuning $\Delta_m$ and squeezing parameter $r$. Here, $g=G=0.1\omega_b$. Other parameters are the same as Fig.~\ref{pptotsqueeze}.}\label{pptotentangle}
\end{figure}
We also simulate the logarithmic negativity (LN) to quantify the photon-phonon squeezing (EPR entanglement)~\cite{entanmeasure,logarithmic,lnent2}, which is defined as 
\begin{equation}\label{ENs}
E_N=\rm {Max}\left\{0,-\frac{1}{2}\rm{ln}\left\{2[\mathcal{P}-(\mathcal{P}^2-4\det V^{\rm eff})^{1/2}]\right\}\right\}.
\end{equation}
$V^{\rm eff}=[V_a, V_{ab}; V^T_{ab}, V_b]$, with $V_a$, $V_b$, and $V_{ab}$ being the $2\times 2$ blocks of $V^{\rm eff}$, and $\mathcal{P}\equiv \det V_a+\det V_b-2\det V_{ab}$. Then, the time-dependent LN can be derived as
\begin{equation}\label{ENveff}
E_N(t)=\rm {Max}\left\{0,-\rm{ln}\left[\eta\left(1-\sqrt{1+\delta'}\right)\right]\right\}.
\end{equation}
where $\eta=V^{\rm eff}_{11}(t)+V^{\rm eff}_{33}(t)$ and $\delta'=\{4[V^{\rm eff}_{13}(t)]^2-4V^{\rm eff}_{11}(t)V^{\rm eff}_{33}(t)\}/\eta^2$. In the unstable dynamical regime $g^2_{eff}>\kappa_a\kappa_b$, using the solutions in Eq.~\eqref{V13t}, one can demonstrate that $\delta'\to 0$ when $t\to\infty$. With the Taylor expansion $\sqrt{1+\delta'}\approx1+\delta'/2$ up to the first order of $\delta'$, one can finally obtain the LN at infinity
\begin{equation}\label{Entinf}
E_N(\infty)=\rm {Max}\left\{0,-\ln [2\Delta X_\phi(\infty)]\right\}.
\end{equation}
When asymptotic stationary two-mode squeezing emerges during unstable dynamics, as indicated by $\Delta X_\phi(\infty)<0.5$, asymptotic stationary bipartite entanglement between the photon and phonon also arises. The LN increases with the value of $\Delta X_\phi(\infty)$ decreases, indicating that stronger two-mode squeezing corresponds to greater entanglement.

The above results obtained by the effective Hamiltonian in Eq.~\eqref{Heff} can be further confirmed by the whole system's dynamics. Similar as Eq.~\eqref{Vijteff}, using the full system Hamiltonian $H$ in Eq.~\eqref{Hamfull}, the dynamics of the whole system CM $V(t)$ is determined by
\begin{equation}\label{Vijt}
\dot{V}(t)=AV(t)+V(t)A^T+D.
\end{equation}
The elements of $V(t)$ are defined as 
\begin{equation}
V_{ij}(t)=\frac{\langle u_i(t)u_j(t)+u_j(t)u_i(t)\rangle}{2}, i,j=1,2\cdot\cdot\cdot 6,
\end{equation}
where $u(t)$ is shown in Eq.~\eqref{utHeisenberg}. The transition matrix $A=i\mathcal{L}+\tilde{A}$, where $\mathcal{L}$ is the superoperator in Eq.~\eqref{Liouvilliansuper} and $\tilde{A}=Diag[-\kappa_a,-\kappa_a,-\kappa_b,-\kappa_b,-e^{2r}\kappa_m,-e^{2r}\kappa_m]$. The magnon decay rate is exponentially enlarged due to the Kerr effect, i.e., $\kappa_m \rightarrow e^{2r}\kappa_m$~\cite{spintripartite}. $D=Diag[\kappa_a(2N_a+1),\kappa_a(2N_a+1),\kappa_b(2N_b+1),\kappa_b(2N_b+1),e^{2r}\kappa_m(2N_m+1),e^{2r}\kappa_m(2N_m+1)]$ is the matrix of noises covariance. Then, the dynamics of $\Delta X(t)$ and $\Delta X_\phi(t)$ can be obtained by numerically calculating the CM $V(t)$, 
\begin{equation}\label{Deltaxwhole}
\begin{aligned}
\Delta X(t)&=\frac{1}{2}[V_{11}(t)+V_{33}(t)+2V_{13}(t)],\\
\Delta X_\phi(t)&=\cos^2\phi V_{11}(t)+\sin^2\phi V_{33}(t)+\sin(2\phi)V_{13}(t).\\
\end{aligned}
\end{equation}
The initial condition is $V(0)=I_6/2$, and $I_6$ is a six-dimensional identity matrix. 

Numerical results are shown in Figs.~\ref{pptotsqueeze}(a) and (b). All of the matrix elements in Fig.~\ref{pptotsqueeze}(a), $V_{11}(t)$ (blue solid line with circles), $V_{33}(t)$ (red dashed line with squares), and $V_{13}(t)$ (black dash-dotted line with diamonds), along with the variances in Fig.~\ref{pptotsqueeze}(b), $\Delta X(t)$ (red dashed line with circles) and $\Delta X_\phi(t)$ (purple dotted line with squares) obtained using Eq.~\eqref{Vijt} do match well with the corresponding results via the effective Hamiltonian~\eqref{Heff}. 

We also numerically analyze the asymptotic stationary LN, $E_N(\infty)$, using the whole system's dynamics to evaluate our protocol. For the whole system, the $E_N$ is defined by replacing the $V^{\rm eff}$ in Eq.~\eqref{ENs} with $V(1:4,1:4)$ in Eq.~\eqref{Vijt}. Besides, we approximate $E_N(\infty)$ by evaluating the value after a sufficient long time, specifically $E_N(\infty)=E_N(2\tau)$, where $\tau$ is the moment when $\Delta X(t)$ in Eq.~\eqref{XYVeff} reaches its minimum, as shown in Fig.~\ref{pptotsqueeze} (b). 

In Fig.~\ref{pptotentangle}(a), we illustrate the $E_N(\infty)$ in the parameter space of coupling strengths $g$ and $G$. One can observe that a high entanglement LN can be obtained when the coupling strengths $g$ and $G$ are increased. The LN $E_N(\infty)$ is greater than $2.5$ for $g, G\ge 0.2\omega_b$. In Fig.~\ref{pptotentangle}(b), we present $E_N(\infty)$ in the parameter space defined by magnon detuning $\Delta_m$ and the squeezing parameter $r$. A low entanglement regime $E_N(\infty)\le0.5$ around $\Delta_m\approx\omega_b$ is observed, and the low-region enhances with increasing $r$, which is attributed to the invalidity of the effective Hamiltonian~\eqref{Heff} constructed in perturbation condition. In the region where $\Delta_m<\omega_b$, the effective coupling strength $g_{\rm eff}$ decreases as $r$ increases [see Fig.~\ref{effgdelta}(e)], which leads to a reduction in LN with increasing $r$. In contrast, when $\Delta_m>\omega_b$, the effective coupling $g_{\rm eff}$ increases as $r$ increases, leading consequently to a enhancement of the LN with increasing $r$. However, under the constraint of the large detuning condition, i.e., $|\Delta_m/\cosh(2r)-\omega_b|\gg g,G$, an even larger $r$ does not always yield better results. A larger $\Delta_m$ allows a broader range of $r$ to achieve a high LN. An optimal parameter space $E_N(\infty)\ge 2$ exists at $\Delta_m\le 3.5\omega_b, r\ge0.15$. From these findings, we conclude that our protocol significantly enhances the asymptotic stationary LN, $E_N$, from the previously reported values of approximately $0.1-0.3$~\cite{mppentang}. It can also exceed the maximum theoretical limit at the stable dynamic condition, i.e., $0.69$~\cite{reservoiroptome}.

\section{Discussion and Conclusion}\label{Secexpercon}
Our protocol is primarily centered on the cavity magnomechanical system, focusing on constructing the magnon-assisted photon-phonon squeezing. In recent experiments~\cite{magnoncavity,kerrcavitymagnon,BackactionMag}, the coupling strength between photon and magnon, $g\sim 1-10$MHz, is about $0.1\omega_b$, where the phonon frequency $\omega_b\sim10-100$MHz. The single-excitation magnon-phonon coupling $g_{mb}$ is related to the volume of the YIG sphere and the biased magnetic field directions. For a YIG sphere with a diameter $100\mu$m, the coupling strength is approximately $g_{mb}\sim 0.1$Hz~\cite{magnoncavity}. The Rabi frequency of driving is defined as $\Omega\equiv\sqrt{\kappa P_d/(\hbar\omega_d)}$~\cite{kerrcavitymagnon,BackactionMag}, where $\kappa$ is the cavity decay rate associated with the driving point, $P_d$ and $\omega_d$ represent the power and frequency of the microwave drive, respectively. In recent experiments~\cite{kerrcavitymagnon}, $\kappa/2\pi\sim 0.1$MHz and $P_d\sim20-30$dBm ($100-1000$mW). They give rise a Rabi frequency $\Omega\sim 10^{14}-10^{15}$Hz, corresponding to $|\langle m\rangle|\approx 10^{6}- 10^7$. The enhanced coupling between magnon and phonon is given by $G\equiv g_{mb}|\langle m\rangle|\sim 0.1\omega_b$. Consequently, the effective coupling strength satisfies $|g_{\rm eff}|\sim 0.01\omega_b$. These values are consistent with the parameters used in Figs.~\ref{eigenLiouvillian} and~\ref{effgdelta}. The magnon Kerr effect can be enhanced by reducing the volume of the YIG sphere~\cite{kerrmagnoncav,theorykerr,kerrmagnonspin}. For specifical parameters $K_m\sim 10$nHz, $\Delta_m=3\omega_b$, and $|\langle m\rangle|=10^7$, the squeezing parameter $r$ roughly equals $0.05$. The decay rates of phonon and photon modes are $\kappa_b\sim10^{-5}-10^{-4}\omega_b$ and $\kappa_a\sim100\kappa_b$, respectively. Thus, the cooperativity $C\equiv g^2_{\rm eff}/\kappa_a\kappa_b \gtrsim 100$. The unstable dynamical regime in our protocol, characterized by $g^2_{\rm eff}>\kappa_a\kappa_b$, can be successfully realized. Besides, the magnon decay rate, $\kappa_m\sim 1\rm{MHz}$, is challenging to reduce further due to intrinsic damping~\cite{quantummagnon}. However, its impact is insignificant as the magnon primarily serves as an interface. At a low temperature of $T\sim10\rm{mK}$, the thermal occupations of photon, magnon, and phonon are respective $N_a\approx0$, $N_m\approx0$, and $N_b\approx10$, consistent with the parameters used in Figs.~\ref{pptotsqueeze} and~\ref{pptotentangle}.

In summary, we have presented a protocol for generating photon-phonon squeezing in the cavity magnomechanics, where the magnon in the YIG sphere is coupled to both microwave photons and the mechanical vibration modes in the same sphere. Our protocol offers significant advantages in terms of controllability within the system, and the Kerr effect of magnon can further enhance the squeezing level in a proper regime. This magnon-assisted protocol relies on the effective two-mode squeezing Hamiltonian for coupling photons and phonons. We apply an interesting method by diagonalizing the whole system's Liouvillian superoperator to numerically confirm the validity of the effective Hamiltonian, which is beneficial to more physics with nonconservative excitations. In the open-quantum-system framework, we derive the process for generating TMSS with the effective Hamiltonian. Our analysis demonstrates that the asymptotic stationary TMSS with high squeezing levels can be achieved even in unstable dynamics. Our work provides an important implementation of TMSS generation in a solid system under realistic noises. It extends the application of cavity magnomechanics as a promising hybrid platform for quantum information processing.

In addition to the cavity magnomechanical system, our protocol can be extended to other quantum systems. For instance, we can utilize a mechanical interface to realize the microwave-optical photon squeezed state~\cite{microillumin, transduction} or create the photon-magnon squeezed state~\cite{cavityoptomag}. Our scheme presents an extendable framework to create TMSS, which will be widely applied in quantum information processing and quantum metrology using bosonic systems.

\section*{Acknowledgments}
We acknowledge financial support from the National Science Foundation of China (Grant No. 12404405) and the Science Foundation of Hebei Normal University of China (Grant No. L2024B10).

\appendix
\begin{widetext}
\section{System linearized Hamiltonian}\label{appalinearHam}
This appendix contributes to deriving the linearized Hamiltonian in Eq.~\eqref{Hamfull}. With respect to the transformation $U(t)=\exp\{i\omega_dta^\dag a+i\omega_dtm^\dag m\}$, the original Hamiltonian in Eq.~\eqref{Hamsysori} turns out to be
\begin{equation}\label{Hamsysden}
\begin{aligned}
H_s&=\Delta_aa^\dag a+\Delta_mm^\dag m-K_mm^\dag mm^\dag m+\omega_bb^\dag b+g_{ma}(a^\dag m+am^\dag)+g_{mb}m^\dag m(b+b^\dag)+\Omega(a^\dag+a),
\end{aligned}
\end{equation}
where $\Delta_a=\omega_a-\omega_d$ and $\Delta_m=\omega_m-\omega_d$. Due to the Heisenberg-Langevin equation, the time evolution of the system operators satisfies
\begin{equation}\label{lange}
\begin{aligned}
\dot{a}&=-(i\Delta_a+\kappa_a)a-ig_{ma
}m-i\Omega+\sqrt{2\kappa_a}a_{in},\\
\dot{m}&=-(i\Delta_m+\kappa_m)m+iK_mmm^\dag m+iK_mm^\dag mm-ig_{ma}a-ig_{mb}m(b+b^\dag)+\sqrt{2\kappa_m}m_{in},\\
\dot{b}&=-(i\omega_b+\kappa_b)b-ig_{mb}m^\dag m+\sqrt{2\kappa_b}b_{in}.\\
\end{aligned}
\end{equation}
where $a_{in}$, $m_{in}$ and $b_{in}$ are the input noise operators for the cavity photon, magnon, and phonon modes, respectively, which are characterized by the covariance functions: $\langle o_{in}(t)o^\dag_{in}(t')\rangle=[N_o+1]\delta(t-t')$ and $\langle o^\dag_{in}(t)o_{in}(t')\rangle=N_o\delta(t-t'), o=a,m,b$, under the Markovian approximation. $N_o=[\exp(\hbar\omega_o/k_BT)-1]^{-1}$ is the mean population of mode $o$ at the thermal equilibrium state. $\kappa_a$, $\kappa_m$, and $\kappa_b$ are the decay rates of the modes $a$, $m$ and $b$, respectively.

Under the condition that the photon mode under strong driving, it has a large amplitude $|\langle a\rangle|\gg1$ at its steady state. Due to the strong photon-magnon dipole-dipole interaction, the magnon mode also has a large amplitude $|\langle m\rangle|\gg1$. This allows us to linearize the system's dynamics around the steady state values by writing the operators $o=\langle o\rangle+\delta o, o=a,m,b$, $\delta_o$ is the operator describing the small quantum fluctuation. The steady values $\langle o\rangle$ satisfy
\begin{equation}\label{steadyvalue}
\begin{aligned}
&-(i\Delta_a+\kappa_a)\langle a\rangle-ig_{ma}\langle m\rangle-i\Omega=0,\\
&-(i\Delta_m+\kappa_m)\langle m\rangle+2iK_m\langle m\rangle|\langle m\rangle^2|-ig_{ma}\langle a\rangle-ig_{mb}\langle m\rangle(\langle b\rangle+\langle b\rangle^*)=0,\\
&-(i\omega_b+\kappa_b)\langle b\rangle-ig_{mb}|\langle m\rangle^2|=0.\\
\end{aligned}
\end{equation}
Then we have
\begin{equation}\label{steadyvalue2}
\begin{aligned}
&\langle a\rangle=-\frac{ig_{ma}\langle m\rangle+i\Omega}{i\Delta_a+\kappa_a},\quad \langle b\rangle=-\frac{ig_{mb}|\langle m\rangle^2|}{i\omega_b+\kappa_b},\\
&-(i\Delta_m+\kappa_m)\langle m\rangle+2i\frac{K_m(\kappa_b^2+\omega^2_b)+g^2_{mb}\omega_b}{\kappa_b^2+\omega^2_b}\langle m\rangle|\langle m\rangle^2|-\frac{g^2_{ma}}{i\Delta_a+\kappa_a}\langle m\rangle-\frac{g_{ma}\Omega}{i\Delta_a+\kappa_a}=0.
\end{aligned}
\end{equation}
When $g_{ma},\kappa_a,\kappa_m\ll|\Delta_a|,|\Delta_m|,\omega_b$ and both $g_{mb}$ and $K_m$ are significantly small, the steady magnitude of magnon mode approximately equals to $|\langle m\rangle|\approx g_{ma}\Omega/|\Delta_m\Delta_a|$.

By substituting the steady values in Eq.~\eqref{steadyvalue2} into the equations in Eq.~\eqref{lange} and ignoring all the high-order terms of fluctuations, the Heisenberg-Langevin equations describing the fluctuation operators $\delta o$ can be written as
\begin{equation}\label{flucoperator}
\begin{aligned}
\dot{\delta a}&=-(i\Delta_a+\kappa_a)\delta a-ig_{ma}\delta m+\sqrt{2\kappa_a}a_{in},\\ 
\dot{\delta m}&=-(i\Delta m+\kappa_m)\delta m+2iK_m|\langle m\rangle^2|\delta m+2iK_m\langle m\rangle^2\delta m^\dag-ig_{ma}\delta a-ig_{mb}\delta m(\langle b\rangle+\langle b\rangle^*)\\
&-ig_{mb}\langle m\rangle(\delta b+\delta b^\dag)+\sqrt{2\kappa_m}m_{in},\\
\dot{\delta b}&=-(i\omega_b+\kappa_b)\delta b-ig_{mb}(\langle m\rangle^*\delta m+\langle m\rangle\delta m^\dag)+\sqrt{2\kappa_b}b_{in}.\\
\end{aligned}
\end{equation}
The corresponding effective linearized Hamiltonian can be described as
\begin{equation}\label{Hamlinear}
\begin{aligned}
H_{\rm lin}&=\Delta_a\delta a^\dag \delta a+\tilde{\Delta}_m\delta m^\dag \delta m+\omega_b\delta b^\dag \delta b-K\delta m^{\dag2}-K^*\delta m^2+g(\delta a \delta m^\dag+\delta a^\dag \delta m)+(G\delta m^\dag+G^*\delta m)(\delta b+\delta b^\dag),
\end{aligned}
\end{equation}
where $\tilde{\Delta}_m=\Delta_m-2|K|$, $K=K_m\langle m\rangle^2$, $g=g_{ma}$, and $G=g_{mb}\langle m\rangle$. We apply the convention $\delta o\rightarrow o, o=a,m,b$ in the following content for simplicity. 

In the rotating frame with the unitary transformation $U(\epsilon)=\exp[\frac{1}{2}(re^{-i\theta}m^2-re^{i\theta}m^{\dag2})]$, the linearized Hamiltonian in Eq.~\eqref{Hamlinear} transforms into
\begin{equation}\label{Hamlinear2}
\begin{aligned}
H_{\rm rot}&=\Delta_aa^\dag a+\omega_bb^\dag b+\tilde{\Delta}_m(m^\dag m\cosh^2r+mm^\dag \sinh^2r+m^{\dag 2}\cosh r\sinh re^{i\theta}+m^2\cosh r\sinh r e^{-i\theta})\\
&-K(m^{\dag2}\cosh^2 r+mm^\dag e^{-i\theta}\cosh r\sinh r+m^\dag m e^{-i\theta}\cosh r\sinh r+m^2e^{-2i\theta}\sinh^2 r)\\
&-K^*(m^2\cosh^2 r+mm^\dag e^{i\theta}\cosh r\sinh r+m^\dag m e^{i\theta}\cosh r\sinh r+m^{\dag 2}e^{2i\theta}\sinh^2 r)\\
&+g\cosh r(am^\dag+a^\dag m)+g\sinh r(ame^{-i\theta}+a^\dag m^\dag e^{i\theta})\\
&+(Gm^\dag\cosh r+Gme^{-i\theta}\sinh r+G^*m\cosh r+G^*m^\dag e^{i\theta}\sinh r)(b+b^\dag).
\end{aligned}
\end{equation}
Setting $\tanh(2r)=2|K|/\tilde{\Delta}_m$ and $K=|K|e^{i\theta}$, the quadratic terms about $m^2$ and $m^{\dag 2}$ can be canceled. Then the Hamiltonian in Eq.~\eqref{Hamlinear2} can be reduced into
\begin{equation}\label{Hameffective}
\begin{aligned}
H&=H_0+V,\quad H_0=\Delta_aa^\dag a+\Delta'_mm^\dag m+\omega_bb^\dag b,\\
V&=g\cosh r(am^\dag+a^\dag m)+g\sinh r(ame^{-i\theta}+a^\dag m^\dag e^{i\theta})+|G|e^r(me^{-i\frac{\theta}{2}}+m^\dag e^{i\frac{\theta}{2}})(b+b^\dag),
\end{aligned}
\end{equation}
where $\Delta'_m=\tilde{\Delta}_m/\cosh(2r)$. The effective magnon coupling strength $G$ can be written as $G=g_{mb}\langle m\rangle=|G|e^{i\theta/2}$ due to $K=K_m\langle m\rangle^2=|K|e^{i\theta}$. It is the linearized Hamiltonian~\eqref{Hamfull} in the main text. For simplicity, we apply the conventions $|G|\to G$ and $\tilde{\Delta}_m\to \Delta_m$ in the main manuscript and the following content.

\section{Effective Hamiltonian for photon-phonon squeezing}\label{appaeffHam}
\begin{figure}[htbp]
\centering
\includegraphics[width=0.8\textwidth]{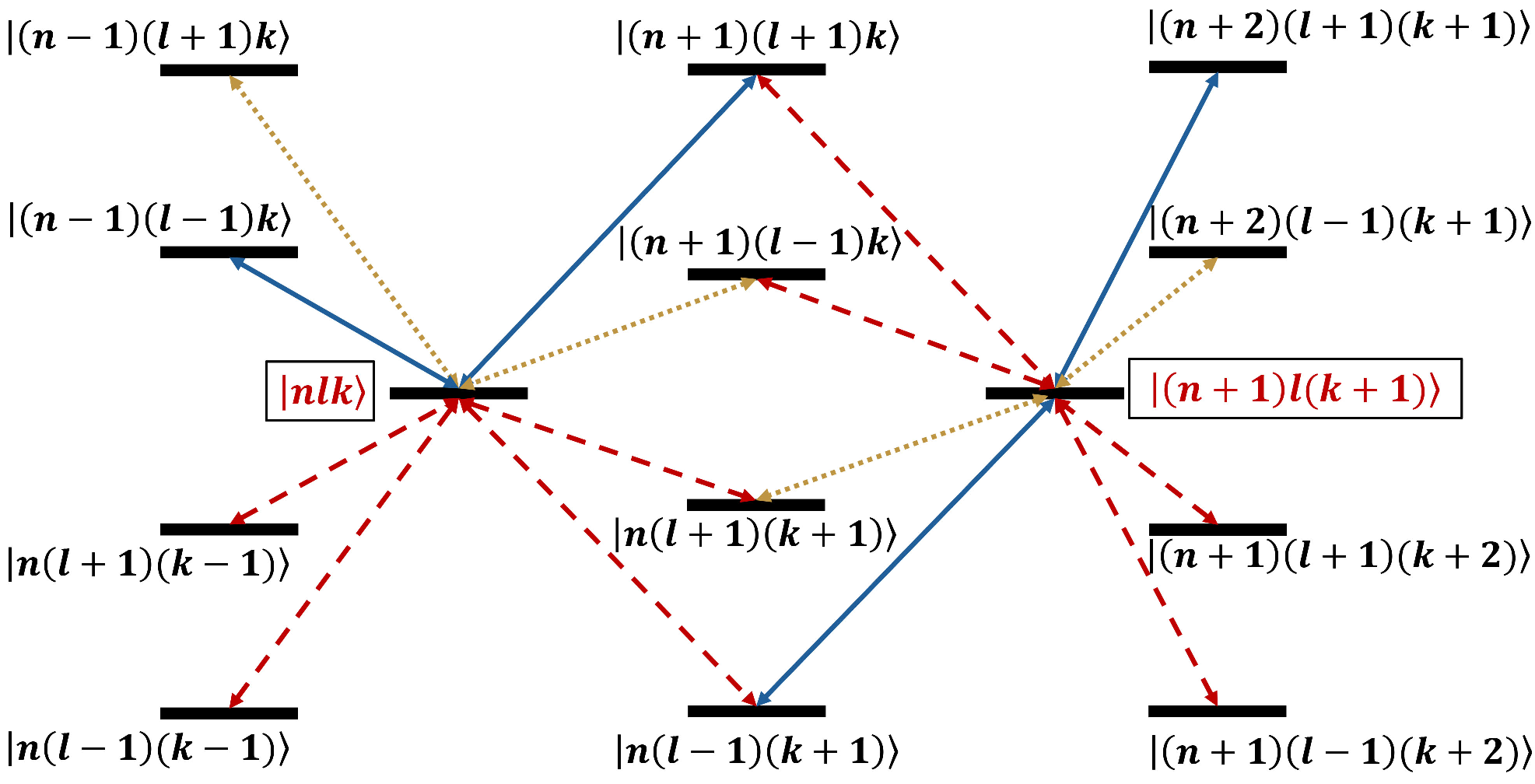}
\caption{All the second-order (leading-order) paths involving arbitrary base pair $|nlk\rangle\equiv|n\rangle_a|l\rangle_m|k\rangle_b$ and $|(n+1)l(k+1)\rangle$. Blue solid (Golden dotted) lines mark the transitions mediated by the counterrotating (rotating) photon-magnon coupling. Red long-dashed lines mark the transitions mediated by magnon-phonon coupling.}\label{eigenpath}
\end{figure}

To realize the photon-phonon squeezing assisted by the magnon mode via the linearized Hamiltonian in Eq.~\eqref{Hameffective}, generally one can extract an effective transition from the near-degenerate subspaces based on the perturbation theory concerning the coupling strengths $g$ and $G$. When the photon detuning frequency $\Delta_a$ is almost opposite the phonon frequency $\omega_b$, and both of them are far resonant from the detuning $\Delta'_m$, i.e., $\Delta_a+\omega_b\approx 0$ and $|\Delta'_m-\Delta_a|,|\Delta'_m-\omega_b| \gg Ge^{r}, g\sinh r,g\cosh r$, it is found that the tensor-product state $|nlk\rangle\equiv |n\rangle_a|l\rangle_m|k\rangle_b$ is near degenerate with $|(n+1)l(k+1)\rangle$. Here the subscripts $a, m, b$ respectively represent the photon, magnon, and phonon modes, and $n, l, k$ indicate their individual Fock states. To the second order, the effective coupling strength or the energy shift between any eigenstates $|i\rangle$ and $|j\rangle$ of the unperturbed Hamiltonian $H_0$ in Eq.~\eqref{Hameffective} is given by~\cite{oneexcite,nonlinear,james}
\begin{equation}\label{secondgeff}
\tilde{g}=\sum_{n\neq i,j}\frac{\langle j|V|n\rangle\langle n|V|i\rangle}{\omega_i-\omega_n},
\end{equation}
where $\omega_n$ is the eigenenergy of state $|n\rangle$, provided the interaction Hamiltonian $V$ is regarded as a perturbation to $H_0$.

A good approximation of the effective Hamiltonian describing the transition between arbitrary base pair $|nlk\rangle$ and $|(n+1)l(k+1)\rangle$ can be analytically obtained using the preceding second-order perturbation theory. It can be expressed in the form
\begin{equation}\label{Heffsubsecond}
H_{\rm eff}=\epsilon_1|nlk\rangle\langle nlk|+(\Delta_a+\omega_b+\epsilon_2)
|(n+1)l(k+1)\rangle\langle (n+1)l(k+1)|+(\tilde{G}|nlk\rangle\langle (n+1)l(k+1)|+\rm{H.c.}),
\end{equation}
where $\epsilon_1$ and $\epsilon_2$ are the energy shifts due to the coupling for the states $|nlk\rangle$ and $|(n+1)l(k+1)\rangle$, respectively. $\tilde{G}$ is the effective coupling strength. These are three coefficients to be determined in this ansatz. We here omit the common unperturbed eigenenergy of two bases $n\Delta_a+l\Delta'_m+k\omega_b$.

We first consider the energy shift $\epsilon_1$ for the state $|nlk\rangle$. Summarizing all the eight paths from $|nlk\rangle$ to $|(n+1)l(k+1)\rangle$ through an intermediate state, as shown in Fig.~\ref{eigenpath}, one can obtain the second-order energy correction $\epsilon_1$ for the $|nlk\rangle$ according to Eq.~\eqref{secondgeff}
\begin{equation}\label{epsilon1second}
\epsilon_1=\frac{(n-l)g^2\cosh^2r}{\Delta_a-\Delta'_m}-\frac{(n+l+1)g^2\sinh^2r}{\Delta_a+\Delta'_m}+\frac{(k-l)G^2e^{2r}}{\omega_b-\Delta'_m}-\frac{(k+l+1)G^2e^{2r}}{\omega_b+\Delta'_m}.
\end{equation}
And in the same way, the energy shift $\epsilon_2$ for the state $|(n+1)l(k+1)\rangle$ is found to be 
\begin{equation}\label{epsilon2second}
\epsilon_2=\frac{(n-l+1)g^2\cosh^2r}{\Delta_a-\Delta'_m}-\frac{(n+l+2)g^2\sinh^2r}{\Delta_a+\Delta'_m}+\frac{(k-l+1)G^2e^{2r}}{\omega_b-\Delta'_m}-\frac{(k+l+2)G^2e^{2r}}{\omega_b+\Delta'_m}.
\end{equation}

An exact resonance between arbitrary $|nlk\rangle$ and $|(n+1)l(k+1)\rangle$ requires that the first two terms in Eq.~\eqref{Heffsubsecond} constitute the identity operator in the relevant subspace. Thus, $\epsilon_1=\Delta_a+\omega_b+\epsilon_2$. Assuming the distance between $\Delta_a$ and $-\omega_b$ is $\delta$, one can have
\begin{equation}\label{deltasecond}
\begin{aligned}
\delta&\equiv\Delta_a+\omega_b=\epsilon_1-\epsilon_2=-\frac{g^2\cosh^2 r}{\Delta_a-\Delta'_m}+\frac{g^2\sinh^2 r}{\Delta_a+\Delta'_m}-\frac{G^2e^{2r}}{\omega_b-\Delta'_m}+\frac{G^2e^{2r}}{\omega_b+\Delta'_m}\\
&=\frac{G^2e^{2r}+g^2\cosh^2r}{\Delta'_m+\omega_b}+\frac{G^2e^{2r}+g^2\sinh^2 r}{\Delta'_m-\omega_b}+\left[\frac{g^2\cosh^2r}{(\Delta'_m+\omega_b)^2}-\frac{g^2\sinh^2r}{(\Delta'_m-\omega_b)^2}\right]\delta+O(\delta^2)\\
&\equiv A+B\delta+O(\delta^2),
\end{aligned}
\end{equation}
where $O(\delta^2)$ represents all the higher orders of $\delta$ from the first order in Taylor expansion. Then $\delta$ is consistently solved as $\delta=A/(1-B)$ up to the second-order correction. Note $B\approx O(g^2/|\Delta'_m-\omega_b|^2)$, so that up to the second order of the coupling strengths $g$ and $G$, we have
\begin{equation}\label{deltareson}
\delta=\frac{2G^2\Delta_m e^{2r}\cosh(2r)+g^2(\Delta_m-\omega_b)\cosh^2(2r)}{\Delta^2_m-\omega^2_b\cosh^2(2r)}
\end{equation} 
via $\Delta'_m=\Delta_m/\cosh(2r)$. Interestingly, $\delta$ is a Fock-state-independent coefficient in comparison to both $\epsilon_1$ and $\epsilon_2$.

Next, we consider the contribution from the four paths connecting $|nlk\rangle$ and $|(n+1)l(k+1)\rangle$ in Fig.~\ref{eigenpath} to their effective coupling strength. By virtue of Eq.~\eqref{secondgeff}, one can have
\begin{equation}\label{geffsub}
\tilde{G}=\sqrt{(n+1)(k+1)}e^{i\frac{\theta}{2}}gG\cosh(2r)\frac{\omega_b\cosh(2r)-\Delta_m e^{2r}}{\Delta^2_m-\omega^2_b\cosh^2(2r)}
\equiv-\sqrt{(n+1)(k+1)}e^{i\frac{\theta}{2}}g_{\rm eff},
\end{equation}
up to the second order of the coupling strengths $g$ and $G$. Eventually, the effective Hamiltonian~\eqref{Heffsubsecond} can be written as
\begin{equation}\label{Heffsub}
H_{\rm eff}=\tilde{G}|nlk\rangle\langle (n+1)l(k+1)|+{\rm{H.c.}}=\left(\tilde{G}|nk\rangle\langle (n+1)(k+1)|+{\rm{H.c.}}\right)\otimes |l\rangle_m\langle l|.
\end{equation}

Discard the magnon mode and extend the Hamiltonian~\eqref{Heffsub} to the whole Hilbert space of photon and phonon, the effective Hamiltonian can be expressed as
\begin{equation}\label{Heffappa}
H_{\rm eff}=g_{\rm eff}(e^{i\frac{\theta}{2}}a^\dag b^\dag+e^{-i\frac{\theta}{2}}ab).
\end{equation}
That is exactly the effective Hamiltonian in Eq.~\eqref{Heff} describing the coupling between photon and phonon. 

\end{widetext}

\bibliographystyle{apsrevlong}
\bibliography{reference}

\end{document}